\documentclass[namedreferences]{SolarPhysics}
\usepackage[optionalrh]{spr-sola-addons} 
\usepackage{graphicx}        
\usepackage{color}           
\usepackage{url}             

\newcommand{\eqref}[1]{(\ref{eqn:#1})}

\begin{document}

\begin{article}

\begin{opening}

\title{Modelling the Global Solar Corona III: Origin of the Hemispheric Pattern of Filaments}

\author{A.R.~\surname{Yeates}\sep
        D.H.~\surname{Mackay}     
       }
\runningauthor{A.R.~Yeates \& D.H.~Mackay}
\runningtitle{Modelling the Global Solar Corona III}

   \institute{School of Mathematics and Statistics, University of St Andrews, Fife, KY16 9SS, Scotland\\
                     email: \url{anthony@mcs.st-and.ac.uk}
             }

\begin{abstract}
We consider the physical origin of the hemispheric pattern of filament chirality on the Sun. Our 3D simulations of the coronal field evolution over a period of 6 months, based on photospheric magnetic measurements, were previously shown to be highly successful at reproducing observed filament chiralities. In this paper we identify and describe the physical mechanisms responsible for this success. The key mechanisms are found to be (1) differential rotation of north-south polarity inversion lines, (2) the shape of bipolar active regions, and (3) evolution of skew over a period of many days. As on the real Sun, the hemispheric pattern in our simulations holds in a statistical sense. Exceptions arise naturally for filaments in certain locations relative to bipolar active regions, or from interactions between a number of active regions. 
\end{abstract}
\end{opening}

\section{Introduction}
Solar filaments (also known as prominences) are classified as having either dextral or sinistral chirality (handedness). The chirality depends on whether the axial magnetic field in the filament points right or left, when seen from the side of the filament with positive magnetic polarity in the photosphere. Observations show that quiescent filaments in the northern hemisphere tend to be dextral, while those in the southern hemisphere tend to be sinistral \cite{rust1967,leroy1983,martin1994,pevtsov2003}. The latter paper found that $80\%$ -- $85\%$ of quiescent filaments follow this hemispheric pattern, which remains the same when the Sun's polar field reverses every 11 years.

Theories for the formation of sheared magnetic fields in filaments are divided between those invoking a sub-surface origin and those requiring purely surface effects. \inlinecite{rust1994} propose that the helical magnetic fields in filaments emerge from the convection zone as already twisted structures, as in the simulations of \inlinecite{gibson2004}. Indeed, observations show that emerging magnetic flux may appear at the solar surface carrying electric current \cite{leka1996}, and in a twisted flux-rope geometry \cite{lites2005}. However, filaments observed so far in such circumstances have been short-lived active region filaments, rather than large stable quiescent filaments. For the latter, the emerging flux rope hypothesis is inconsistent with the tendency for filaments to form between multiple active regions, rather than in centres of flux emergence activity \cite{mackay2008}.

The alternative origin of sheared fields in filament channels is surface motions; we know that after active regions emerge they are advected by the large-scale surface motions of differential rotation and meridional flow, and the flux is dispersed by the action of supergranular convection. The basic principle for formation of axial magnetic fields along polarity inversion lines (PILs) was proposed by \inlinecite{vanballegooijen1989}, based on flux cancellation at the PIL following surface shearing. More recent models have considered the interaction of two bipolar active regions, where the filament forms above a PIL between them \cite{martens2001,litvinenko2005,welsch2005,mackay3}. The latter paper allowed the bipoles to have a non-zero magnetic helicity (twist), effectively including the emergence of twisted fields from below the surface. As the helicity of active regions is observed to follow a hemispheric pattern \cite{pevtsov1995}, this hemispheric pattern of twist was suggested to be a contributory factor to the hemispheric pattern in filament chirality.

The work of \inlinecite{mackay3}, who considered the basic interaction between a pair of magnetic bipoles, is the motivation for the present study. We have constructed a new version of the numerical model for the evolution of the global solar corona over multiple Carrington rotations, based on observations of photospheric magnetic field distributions. The model has been described in detail in \citeauthor{yeates2007a} (\citeyear{yeates2007a}, hereafter ``Paper 1'') and \citeauthor{yeates2008a} (\citeyear{yeates2008a}, hereafter ``Paper 2''), and is summarised very briefly in Section \ref{sec:model} below. In the second paper, we compared the chiralities of 109 observed filaments with the skew angle of our simulated coronal field at the observed locations of the filaments. We found that the model was highly successful in reproducing the chirality of observed filaments, including exceptions to the hemispheric pattern. In this paper we consider the specific mechanisms responsible for the chirality in our simulation. We have identified eight key mechanisms which produce skewed coronal fields above PILs, and each mechanism is described in Section \ref{sec:mech} before considering their relative importance in Section \ref{sec:results}. We conclude in Section \ref{sec:conclusion} with a summary of the cause of the hemispheric pattern in our simulation.

\section{Overview of the Model} \label{sec:model}
\begin{figure}
\centerline{
\includegraphics[width=1.0\textwidth,clip=]{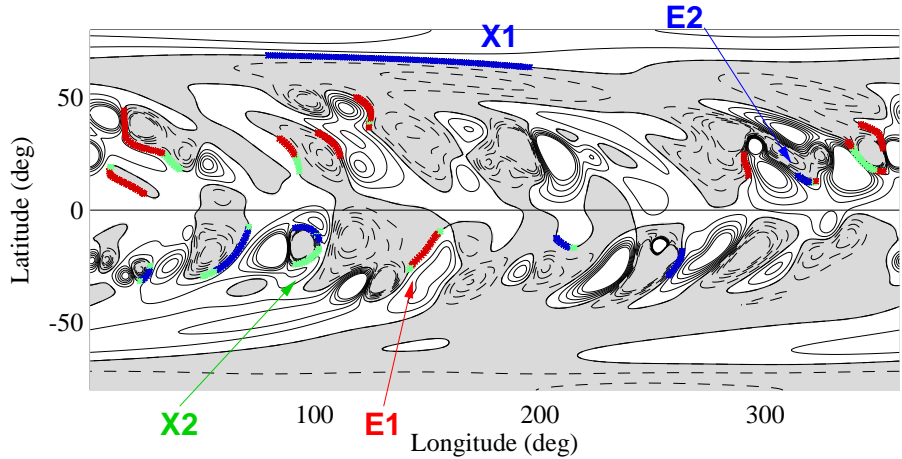}
}
\caption{Surface radial field on day 229, and chirality of simulated filaments during Carrington Rotation 1952. Shading shows the sign of radial magnetic field on the photosphere (white for positive, grey for negative), and thin lines show contours of the same field (solid for positive, dashed for negative). Coloured lines show simulated skew at the locations of observed filaments in the comparison (red for dextral, blue for sinistral, and green for weak skew). }
  \label{fig:fils1952}
\end{figure}

Our model for the 3D coronal field evolution is based on that of \inlinecite{vanballegooijen2000}, and has been described in Papers 1 and 2. The magnetic field in the corona evolves over time in response to (1) the emergence of new magnetic bipoles from below the photosphere, and (2) advection by large-scale motions on the lower (photospheric) boundary. The newly-emerging bipoles take a simple mathematical form and, in order that the simulation remains accurate to the observed fields on the Sun, the bipole properties (latitude, longitude, size, and magnetic flux) are determined from observations. Figure \ref{fig:fils1952} shows a snapshot of the photospheric field on day of year 229, mid-way through the 6-month simulation. Visible are newly-emerged bipoles at low latitudes, and also older regions which have been sheared by differential rotation, have been transported poleward by the meridional circulation, and have interacted with the surrounding field by surface diffusion (representing the cumulative effect of supergranular convection) and reconnection.

The coronal field associated with the photospheric field is evolved using the magnetofrictional relaxation technique, effectively through a continuous sequence of nonlinear force-free equilibria that allows the build up of sheared fields and magnetic helicity. The helicity in our model originates from three distinct sources (illustrated in \opencite{yeates2008b}): (1) new bipoles emerge twisted, (2) currents develop at the interface between old and new flux regions, and (3) surface motions shear the coronal magnetic field. Note that the subsurface origin of emerging helicity is outside the scope of our simulations, and is still uncertain. Theories include the interaction of rising flux tubes with the large-scale poloidal field \cite{choudhuri2004}, with helical turbulence (the $\Sigma$-effect, \opencite{longcope1998}), or with the Coriolis force \cite{holder2004}. Observations of current helicity ({\it e.g.} \opencite{pevtsov1995}; \opencite{hagino2004}) and of X-ray sigmoids \cite{canfield1999} show that the majority twist of active regions is negative in the northern hemisphere and positive in the southern hemisphere. As described in Paper 2, we have run simulations where the sign of emerging-bipole twist ($\beta$) matches this majority observed sign in each hemisphere, and simulations where it takes the opposite sign. The filament hemispheric pattern is best reproduced for the majority observed sign of $\beta$.

The coloured lines in Figure \ref{fig:fils1952} show the skew type ({\it i.e.}, dextral, sinistral, or weak) of our simulated coronal field at the locations which were compared with observed filaments during a particular 27-day period. This shows results from the simulation run with correct hemispheric sign of bipole twist. The majority of these filament locations match the observed chirality, including two exceptions to the hemispheric pattern, labelled E1 and E2 on the figure. The only filaments where our simulation produces the wrong chirality are one on the polar crown (X1), and another at lower latitude (X2) which is only weakly skewed. In this paper we address the question of why our simulation is so successful at reproducing the observed chiralities.

\section{Physical Mechanisms Producing Skew} \label{sec:mech}
\begin{figure}
\centerline{
\includegraphics[width=1.0\textwidth,clip=]{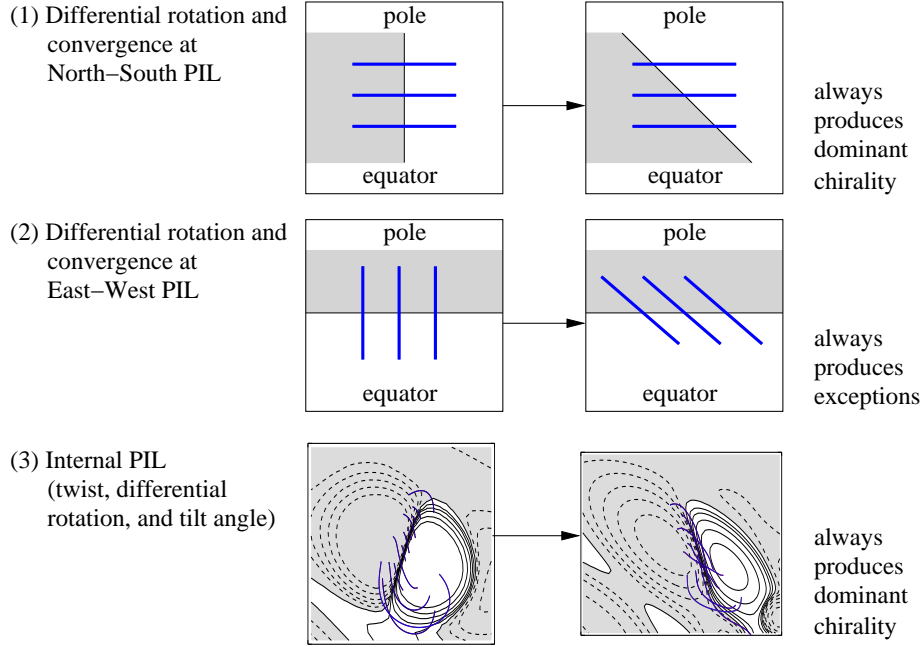}
}
\caption{Mechanisms which always produce the same type of skew ({\it i.e.}, dextral or sinistral) in each hemisphere. In each example, shading shows the sign of radial field $B_r$ on the photosphere (white for positive, grey for negative), thin lines show contours of the same field (solid for positive, dashed for negative), and thick blue lines show selected coronal field lines.}
  \label{fig:mechs1}
\end{figure}
\begin{figure}
\centerline{
\includegraphics[width=1.0\textwidth,clip=]{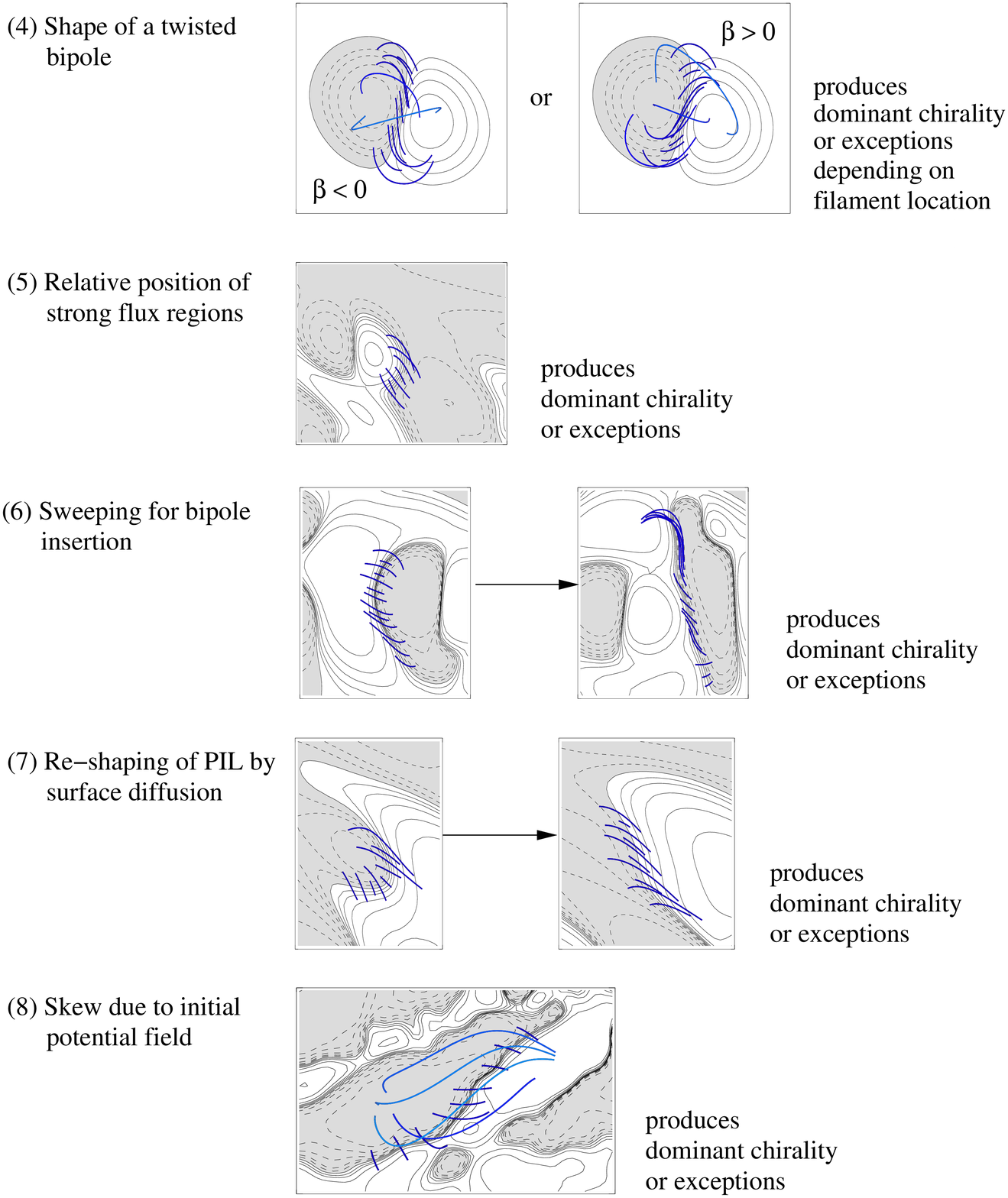}
}
\caption{Mechanisms which can produce either dextral or sinistral skew depending on individual circumstances. In each example, shading shows the sign of radial field $B_r$ on the photosphere (white for positive, grey for negative), thin lines show contours of the same field (solid for positive, dashed for negative), and thick blue lines show selected coronal field lines.}
  \label{fig:mechs2}
\end{figure}

Among our 109 simulated filaments we have identified eight different mechanisms which produce skewed coronal magnetic fields above PILs. Three of these mechanisms always produce the same type of skew ({\it i.e.}, dextral or sinistral) in each hemisphere, and are illustrated in Figure \ref{fig:mechs1}. The other five mechanisms may produce either type of skew depending on individual circumstances, and are illustrated in Figure \ref{fig:mechs2}. Note that changing the polarity of the magnetic field in each figure does not affect the skew type. We describe each mechanism in more detail below.

\subsection{Differential Rotation} \label{sec:diffrot}
The most conceptually simple way to produce a skewed coronal field is through differential rotation of a pre-existing field. The type of skew produced depends on the orientation of the initial field and the PIL. Consider first a coronal field oriented in the east-west direction, over a north-south PIL. This is mechanism 1, illustrated by a cartoon in Figure \ref{fig:mechs1} (for the northern hemisphere). Here differential rotation has little effect on the higher coronal field lines, as both footpoints are at the same latitude. However, the photospheric PIL itself is sheared and so the coronal field develops a skew relative to it. Because the equator rotates faster than the poles, this skew will always be dextral in the northern hemisphere and sinistral in the southern hemisphere, in accordance with the hemispheric pattern. This effect of differential rotation of a north-south PIL was noted by \inlinecite{zirker1997}, but earlier authors had dismissed differential rotation as a potential source of the hemispheric pattern because they had considered a north-south field across an east-west PIL ({\it e.g.}, \opencite{vanballegooijen1990}; \opencite{rust1994}; \opencite{priest1996}). This alternative orientation of field and PIL is mechanism 2 in Figure \ref{fig:mechs1}. In this case the PIL is unaffected by the differential rotation, but the footpoints of the coronal field lines are shifted, resulting in the wrong type of skew, sinistral in the northern hemisphere and dextral in the southern hemisphere. Thus mechanism 2 always acts to produce exceptions to the hemispheric pattern.

\begin{figure}
\centerline{
\includegraphics[width=0.55\textwidth,clip=]{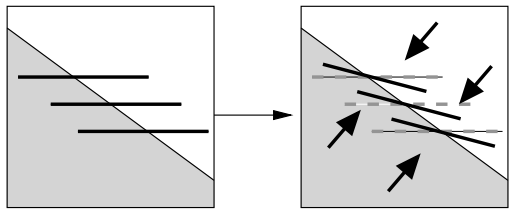}
}
\caption{Schematic showing how surface diffusion converges field line foot-points towards a PIL, leading to an increase in the skew angle.}
  \label{fig:convergence}
\end{figure}

We remark here that the skew developed by differential rotation in mechanisms 1 and 2 (or indeed by any of the other mechanisms described in this paper) is enhanced over time by convergence of flux towards the PIL, owing to supergranular diffusion. This convergence, sketched in Figure \ref{fig:convergence}, moves the footpoints of coronal field lines towards the PIL irrespective of its north-south or east-west orientation, thus enhancing any existing skew, although not producing it from scratch. The process is well illustrated by the H$\alpha$ sequence in Figure 4 of \inlinecite{wang2007}. Note that our simulations do not include any extra converging flow towards the PIL over and above that due to diffusion; although such localised flows have been invoked in some models of filament formation ({\it e.g.}, \opencite{kuijpers1997}) they are not supported by observations \cite{hindman2006}.

\subsection{Filaments on the Internal PIL of a Bipole} \label{sec:internal}
In our simulation there are 32 filaments (out of 109) which form internal to a single active region. The skew of these filaments shows a characteristic behaviour (mechanism 3, Figure \ref{fig:mechs1}). Initially the skew depends only on the sign of $\beta$ -- the bipole twist for the region, but after 1 to 2 Carrington rotations the effects of differential rotation lead to dextral skew in the northern hemisphere and sinistral in the southern hemisphere, irrespective of the sign of $\beta$. This is essentially mechanism 1 in action, but it is assisted by the tilt angle of the bipole, which enhances the latitudinal shearing. So for the majority hemispheric sign of $\beta$ such internal filaments will always follow the hemispheric pattern, and even for the minority sign of $\beta$ the hemispheric pattern will be recovered after 1 to 2 Carrington rotations (30-60 days).

\subsection{Filaments on the External PIL of a Bipole} \label{sec:edge}
Many filaments occur along the outside edge (periphery) of a single bipolar region, and in the simulation their skew is directly affected by the shape of field lines in the bipole (mechanism 4, Figure \ref{fig:mechs2}). This idea is supported by the recent observations of filament formation by \inlinecite{wang2007}. They show an example where H$\alpha$ fibrils on the outside edge of a newly-emerged active region become rapidly aligned with the horizontal field component of the bipole, which lies primarily along the bipole's external PIL (see their Figure 5).

\begin{figure}
\centerline{
\includegraphics[width=1.0\textwidth,clip=]{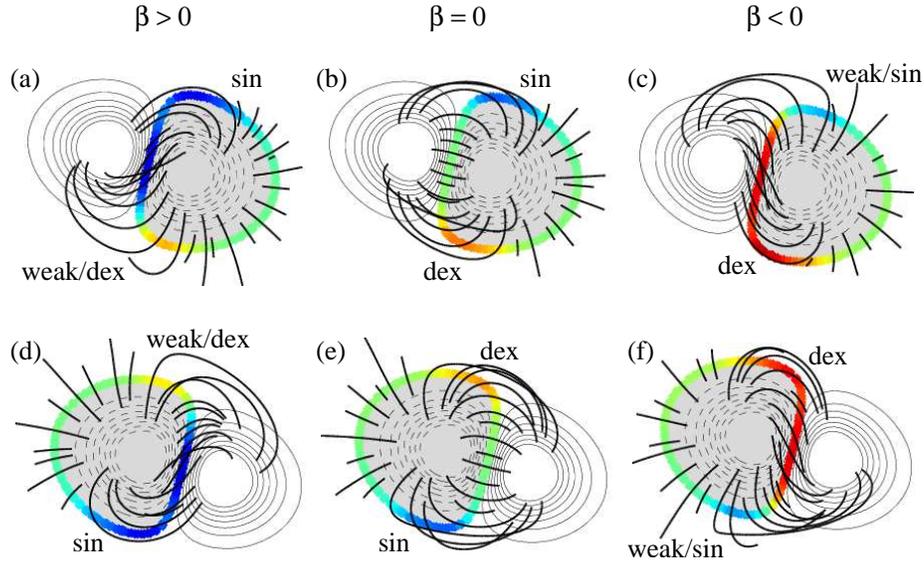}
}
\caption{Shape of the field lines around the PIL of a bipole in a background field. Colours show the skew angle along the PIL for each possible sign combination of bipole twist and background field polarity. The colour table runs from red (dextral skew) through green (weak skew) to blue (sinistral skew). In panels a and d (left column) the bipole has positive twist $\beta$, in panels b and e (centre column) the bipole is untwisted ($\beta=0$), and in panels c and f (right column) the bipole has negative twist. In panels a, b, and c (top row) the following polarity of the bipole matches that of the background field, and in panels d, e, and f (bottom row) the leading polarity matches the background. The shading shows polarity of radial magnetic field on the photosphere (white positive, grey negative), and thin lines show contours of the same field. Thick lines are selected coronal magnetic field lines crossing the PIL.
}
  \label{fig:edge}
\end{figure}

The skew of such a filament depends primarily on where along the bipole boundary the filament lies, and on the polarity of the background field relative to the bipole. The field lines are most strongly skewed at the north and south ends of the bipole, with weaker skew on the middle part of the external PIL (at least until twist has diffused out from the bipole centre). This is seen in Figure \ref{fig:edge}, which shows the shape of the bipole for the six possible sign combinations of background field and bipole twist $\beta$. The type of skew (dextral, sinistral, or weak) is indicated by the colouring along the PIL in each case. Notice that changing the sign of $\beta$ for the bipole does not change the type of skew on the external PIL, but merely the strength of skew. In our simulation there are 31 filaments occuring at either the north or south end of such an external PIL, and the skew pattern in Figure \ref{fig:edge} is seen to hold for the majority (24) of such filaments. 

How does this relate to the hemispheric pattern? In the northern hemisphere, the observed hemispheric pattern of helicity gives $\beta<0$. If the background field has the following polarity then Figure \ref{fig:edge}(c) shows that a filament occuring at the north end of the external PIL will have weak or slightly sinistral skew, whereas one occuring at the south end will have dextral skew. If the background field has the leading polarity, then Figure \ref{fig:edge}(f) shows that a filament at the north end will be dextral, whereas at the south end the field is weakly or slightly sinistrally skewed. Overall, then, there will be more dextral skew produced by this mechanism than sinistral skew in the northern hemisphere. A similar argument shows that in the southern hemisphere, where $\beta>0$,  there will be more sinistral skew produced. Thus filaments on the external PIL of a bipole will tend to favour the hemispheric pattern. Importantly, however, the bipole shape can produce exceptions to the hemispheric pattern for filaments occuring at either the north or south end of a bipole, depending on the relative polarity of the background field.

\subsection{Multiple Bipole Interactions} \label{sec:multiple}

The shape of a single region has a strong influence on the skew of the field above its external PIL, as described in Section \ref{sec:edge}. However, in a minority of cases a non-uniform background field modifies how the bipole connects to other nearby bipoles or regions of strong flux. This is mechanism 5 in Figure \ref{fig:mechs2}, where the example shows a bipole which has emerged to the north of a strong negative region. Rather than connecting directly across the PIL, the field lines from the positive leading polarity have been diverted southward to connect with the stronger region. Thus a part of the PIL which would normally be expected from Figure \ref{fig:edge} to have weak skew actually has strong dextral skew - in fact this was the location of an observed dextral filament. Of course, this mechanism may also produce sinistral skew, as would have been the case had the strong background region been located to the \emph{north} of the new bipole.

In addition to the shape of a bipole being modified by the pre-existing field, the skew of the pre-existing field may also be modified by a new bipole emergence, through the dynamic effect of sweeping (mechanism 6, Figure \ref{fig:mechs2}). The technique of sweeping before the insertion of a new bipole was described in Papers 1 and 2, and results in a more realistic end configuration for newly-emerged regions. In the example shown here in Figure \ref{fig:mechs2}, a new region emerges to the left of a pre-existing PIL with slight dextral skew. The pre-existing positive field has been swept out of the insertion region, resulting in (1) convergence of footpoints towards the pre-existing PIL, and (2) lengthening of field lines across this PIL because the positive footpoints are displaced northward. The result is a highly skewed dextral flux rope structure along the PIL. Again, this mechanism could also produce the opposite sign of skew, and is also capable of introducing skew at a PIL where there was previously none. Although our sweeping mechanism is artificial, we might expect a similar effect on pre-existing coronal field even from a gradual flux emergence.

It was described in Section \ref{sec:diffrot} how differential rotation is the primary surface transport effect that develops skewed coronal fields from scratch, and that surface diffusion acts primarily to enhance existing skew. In fact, in a small number of cases, diffusion does cause skew to develop from scratch, and this is mechanism 7 of Figure \ref{fig:mechs2}. The necessary ingredients are a PIL with initially complex shape. Diffusion inevitably smooths out the PIL, but the coronal field is not necessarily smoothed to the same extent. Thus the final PIL shape may end up sheared with respect to the overlying coronal field lines. \inlinecite{wang2007} show an observed sequence in their Figures 2 and 3 where this mechanism appears to occur.

\subsection{Skew from the Initial Potential Field} \label{sec:potential}

Finally, mechanism 8 in Figure \ref{fig:mechs2} relates to the initial condition of the simulation, which is a potential field extrapolation. Although this field has the minimum energy for the boundary conditions, and so the majority of field lines are unsheared and pass over PILs at $90^\circ$, this is not necessarily always the case. The example in Figure \ref{fig:mechs2} shows an example of how skew may arise in such a potential field. Although the lower field lines are not skewed, higher coronal field lines must link regions of strong flux, so in this case are skewed with respect to the PIL because of the relative locations of the two strong-flux regions. On the real Sun---with no initial potential field---it is unclear whether these overlying field lines would follow the same path.

\section{Relative Importance of Different Mechanisms} \label{sec:results}
To assess the role of each of the eight mechanisms in producing the hemispheric pattern, we present here the results of a detailed analysis of the 109 filaments which were compared to observations in Paper 2. Table \ref{tab:results} shows the number of filaments in relation to each mechanism. Note that the mechanisms are not mutually exclusive; there are 56 filaments with two or more mechanisms associated, and 12 with four or more. The table also breaks various statistics down by mechanism, and these are described in this section. The bottom row shows the same statistics for the full sample of 109 filaments.

\begin{table}
\caption{Classification of filaments by skew mechanism (as defined in Figures \ref{fig:mechs1} and \ref{fig:mechs2}). Simulation results presented here are for run 4 (bipole twist $|\beta|=0.4$, with observed hemispheric sign of helicity), except for the fifth column which compares run 4 with run 1 ($|\beta|=0.2$, opposite sign of helicity).}
\label{tab:results}
\begin{tabular}{llllll} \hline
Mechanism & Number    & \% of filaments & \% of filaments & \% of filaments      & Mean\\
          & of        & following       & matching        & where skew           & number of\\
          & filaments & hemispheric     & observed        & changes with $\beta$ & associated\\
          &           & pattern         & chirality       &                      & bipoles\\
\hline
(1) & 32 & 100.0 & 96.9 & 12.5 & 2.06\\
(2) & 12 &  16.7 & 58.3 &  8.3 & 0.58\\
(3) & 37 & 100.0 & 94.6 & 51.4 & 1.30\\
(4) & 55 &  72.7 & 89.1 & 32.7 & 2.24\\
(5) & 23 &  82.6 & 91.3 & 26.1 & 2.70\\
(6) & 19 &  78.9 & 94.7 & 31.6 & 3.89\\
(7) &  8 &  87.5 & 87.5 &  0.0 & 1.38\\
(8) & 20 &  85.0 & 95.0 & 15.0 & 1.85\\
&&&&&\\
Overall &109 &  78.0 & 86.2 & 32.1 & 1.58\\ 
\hline
\end{tabular}
\end{table}

The most common mechanism found is mechanism 4 (bipole shape), followed by mechanisms 3 (internal filaments) and 1 (differential rotation at a north-south PIL). By contrast, differential rotation at an east-west PIL (mechanism 2) is relatively rare, as is the creation of skew by surface diffusion of a complex PIL (mechanism 7). Note that mechanism 2 may be slightly more common on the real Sun since our sample of 109 filaments is biased against those on the polar crown, due to projection effects hindering chirality observations.

The third column of Table \ref{tab:results} shows the percentage of simulated filaments following the hemispheric pattern. This is for the simulation run with $|\beta|=0.4$ and the observed hemispheric sign of bipole helicity (run 4 in Paper 2). This was the run which best agreed with the observed filament chiralities. As indicated in Section \ref{sec:mech}, mechanisms 1 and 3 always produce skew following the hemispheric pattern, and mechanism 2 produces exceptions. The percentage for mechanism 2 is above zero because it was not the dominant mechanism in all cases where there were multiple factors at play. As expected, mechanisms 4 to 8 produce some filaments following the hemispheric pattern and some not. However, we see that the percentages of filaments following the pattern are in the range 70\%-88\%, which explains the overall result. The lowest percentage for these mechanisms is mechanism 4 (bipole shape), and it was shown in Section \ref{sec:edge} how this can naturally be responsible for exceptions to the hemispheric pattern. The question arises as to how mechanisms 5, 6, and 7 preferentially give the correct pattern, even though they can produce both types of skew. The answer is that they are usually not the only mechanism acting. They tend to enhance existing skew rather than generate completely new skew from scratch. Thus mechanisms 1, 3, and 4 (differential rotation and bipole shape) are more fundamental.

Column 4 shows how the simulated filaments involving each mechanism match with the observed chirality at those locations. The agreement is generally good (above 85\%). Filaments with mechanism 2 (east-west PIL) are least likely to be correct, supporting the idea that this mechanism is not responsible for the chirality of filaments on the Sun. In the simulation this mechanism is associated primarily with high-latitude filaments forming from the initial potential field, whereas in real-life we expect that the correct hemispheric sign of skew will be transported polewards from lower latitudes over longer periods of time.

The fifth column of Table \ref{tab:results} shows the relative importance of emerging active-region helicity for the filament hemispheric pattern. It compares a run with observed hemispheric sign of $\beta$ (run 4 in Paper 2) against a run with the opposite sign of $\beta$ (run 1). A higher percentage means that there are more filaments where changing the bipole twist $\beta$ has an effect on the resulting chirality. From the column we see that $\beta$ has the most influence on internal filaments (mechanism 3), and on filaments formed by mechanisms 4, 5, and 6. These mechanisms are all associated with the emergence of new bipolar regions. However, even for mechanism 3 the percentage is only 51.4\%, because the effect of emerging twist is over-ridden after 1 to 2 Carrington rotations by differential rotation (as described in Section \ref{sec:internal}). This shows the importance of carrying out continuous long-term simulations rather than individual extrapolations.

For mechanisms 2, 7, and 8 the emerging twist has less influence, as these are largely associated with the initial potential field. Overall only 32\% of filaments change their skew type when the hemispheric sign of emerging bipole twist is reversed, and even if filaments related to the initial field are discounted, the percentage will still be below fifty. Thus the emergence of twisted magnetic fields is not the most important factor in producing the hemispheric pattern, especially for the majority of filaments which form between multiple bipolar regions. This explains why our simulation with the wrong sign of $\beta$ (run 1 in Paper 2) still produced the correct chirality for over 60\% of filaments.

Each filament in the sample was associated with a subset of the 119 emerging bipoles in the simulation. These were selected as being those which influenced the skew above the PIL in some way, rather than simply being located nearby. The final column of Table \ref{tab:results} gives the mean number of bipoles associated with each filament, again broken down by mechanism. Not surprisingly, mechanisms 5 (relative positions of flux) and 6 (sweeping) are associated with the most bipoles, while mechanisms 2 and 3 have the fewest. In the case of mechanism 2 (east-west PIL) this is because such locations are at high latitudes and created from the initial potential field only. The overall mean is 1.58 regions, demonstrating that most filaments are associated with more than one active region. This supports the result of \inlinecite{mackay2008}.

Finally, consider how Table 1 would differ if $\beta=0$, {\it i.e.} if the bipoles emerge untwisted (run 2 in Paper 2). The same mechanisms generally occur in that run, except that mechanism 3 (internal PIL) generates skew more slowly because of the lack of initial twist. Indeed, this explains 7 out of the 10 filaments found to be weakly skewed in run 2 (see Table 3 in Paper 2). As a consequence, the overall percentages of filaments following the hemispheric pattern and matching the observed chirality would decrease, but only slightly. Figures 5(b) and (e) show that, even for an untwisted bipole, there is still skew at either end of the external PIL, so that mechanism 4 (bipole shape) and other mechanisms still occur.

\section{Conclusion} \label{sec:conclusion}
In this paper we have presented the results of a detailed analysis of how skew is formed at the filament locations in our 3D simulation of the global solar coronal evolution over 6 months (see Paper 2 for details of the simulation). The simulation produced the correct chirality for up to 96\% of the 109 filaments compared with observations, an unprecedented result. We have identified eight key mechanisms that produce skewed coronal fields above PILs, and these are summarised in Figures \ref{fig:mechs1} and \ref{fig:mechs2}. The relative importance of these mechanisms has been assessed and we are now in a position to explain why the simulation produces the correct hemispheric pattern. With the exception of mechanism 8 (connected to our initial potential field), we suggest that all of these mechanisms occur in some form on the real Sun, so that our explanation for the simulated hemispheric pattern should offer some insight into the origin of the real hemispheric pattern.

There is not one single factor responsible for the pattern, but rather a number of factors depending on the location of each individual filament and the nature of the PIL. The pattern arises as a statistical result because of the relative frequency of these mechanisms. The most important single influence is the bipolar shape of active regions, which is a well-observed feature, except for some activity complexes. The location where a filament forms around or inside such a bipole has a key influence on its skew type. We have shown that on average the hemispheric pattern is upheld for filaments forming either on the internal PIL of a bipole, or around its external edge. The second key mechanism is differential rotation, but acting on a north-south PIL rather than an east-west PIL, as the latter would tend to produce exceptions to the hemispheric pattern. For a north-south PIL, differential rotation leads naturally to dextral skew in the northern hemisphere and sinistral in the southern hemisphere. Such north-south PILs occur frequently both within individual bipoles and between neighbouring active regions, the locations where many filaments are observed to form.

Once an initial skew is produced by one of the fundamental mechanisms described above, it is often enhanced by the action of surface (supergranular) diffusion, which converges the footpoints of coronal field lines towards the PIL, increasing the skew angle. Depending on individual circumstances, the skew may also be either enhanced or weakened by one of the remaining mechanisms in our list, such as the distribution of strong flux regions in the background field, or nearby bipole insertions. In some cases the lift-off of a twisted magnetic flux rope (as in \opencite{mackay4}) was found to remove skew, though typically only from higher field lines. It was found that the original source of skew at each filament location could develop many days before the filament's date of observation (over 100 days for many filaments in this simulation). Over this time a number of emerging active regions could affect the skew of the field through the various mechanisms described. This highlights the importance of simulating the long-term coronal evolution, rather than doing single-time extrapolations.

A fundamental problem which must be addressed by any theory hoping to explain the hemispheric pattern is how exceptions to the pattern may be formed. The results of our study demonstrate that in our model the pattern holds in a statistical sense, and that exceptions are a natural feature. Out of the 109 filaments in our sample, there were 16 with observed chirality opposite to the hemispheric pattern. The majority of these formed on the outside edge of a bipolar region, and six are simply explained by the shape of the bipolar field, as explained in Section \ref{sec:edge}. Three others involve both the shape of the bipole and its relation to the surrounding field (mechanisms 4, 5, and 6 in Figure \ref{fig:mechs2}), giving examples of how such interactions can produce exceptions to the hemispheric pattern in some circumstances. Of the remaining filaments, four involve flux from the initial potential field, where our simulation is not expected to be realistic, and two are internal to individual bipoles. We assume here that on the real Sun these two active regions must have had the minority sign of helicity for their hemisphere; in the simulation run with opposite sign of bipole twist we reproduce the observed chiralities for these filaments.

In summary, we have demonstrated how the hemispheric pattern of filament chirality on the Sun may be explained by surface flux transport and the emergence of twisted active regions. The key elements are (1) differential rotation at north-south PILs and (2) the shape of bipolar active regions, along with an evolution over periods of many days. However, there were many complex examples where interactions between a number of active regions became important. These interactions involved multiple bipoles of different sizes, relative locations, and dates of emergence, including but not limited to the particular two-bipole configurations suggested by \inlinecite{martens2001}. Our simulations support their view (and that of \opencite{zirker1997}) that the skew above PILs on the Sun develops over a period of months, through interaction and merging of multiple active regions. There are observed cases where strong skew appears to develop within a day or two ({\it e.g.} \opencite{gaizauskas1997}), and this sometimes occurs in the simulation following new bipole emergence. However, by comparing runs with opposite signs of emerging bipole helicity, we find that the sign of $\beta$ affects the chirality of only about one third of filaments, conflicting with simple emerging flux rope explanations for the hemispheric pattern \cite{rust1994}.

Finally, we note that our 6-month simulation does not produce the correct chirality for high-latitude filaments, including those on the polar crown. In the present simulation, differential rotation of the initially potential field produces the opposite skew across these primarily east-west PILs. This opposite sign of skew at high latitudes is also reflected in a sign-reversal of current helicity at these latitudes, which has been found to persist for over two years in longer simulations \cite{yeates2008b}. Interestingly, observations of coronal X-ray arcades by \inlinecite{mcallister1998} found their skew to agree with our present simulations, and to be opposite to their underlying filament chiralities. Future simulations will need to consider whether poleward helicity transport over the full solar cycle is enough to counteract the effect of differential rotation on the chirality of high-latitude filaments.

\begin{acks}
The authors acknowledge financial support from the UK Science and Technology Facilities Council (STFC). We thank A.~A. van Ballegooijen and an anonymous referee for suggesting improvements to the paper.
\end{acks}

\end{article} 
\end{document}